\documentclass[aps,prx,superscriptaddress, onecolumn, amsmath,amssymb,floatfix,longbibliography,11pt]
{revtex4-2} %two columns
\usepackage{bm, comment, graphicx, here, amsmath, color, mathrsfs}
\usepackage[normalem]{ulem}
\usepackage[colorlinks=True, allcolors=blue]{hyperref}
\usepackage{balance, setspace, float, titlesec, array}
\usepackage{ulem}
\titlespacing*{\section}{0pt}{1ex}{0ex}
\titlespacing*{\subsection}{0pt}{1ex}{0ex}
\usepackage{hyperref}

\newcommand{\blue}[1]{ \textcolor{blue}{#1} }

\usepackage{soul}
\soulregister\cite7
\soulregister\ref7
\soulregister\onlinecite7

\begin{document}
\title{PyMatterSim: a Python Data Analysis Library for Computer Simulations of Materials Science, Physics, Chemistry, and Beyond}
\author{Y.-C. Hu\footnote{Email: ychu0213@gmail.com}}
\affiliation{Songshan Lake Materials Laboratory, Dongguan, China}
\author{J. Tian}
\affiliation{Lawrence Berkeley National Laboratory, Berkeley, CA 94720, USA}

\date{\today}

\begin{abstract}
Computer simulation has become one of the most important tools in scientific research in many disciplines. Benefiting from the dynamical trajectories regulated by versatile interatomic interactions, various material properties can be quantitatively characterized at the atomic scale. This greatly deepens our understanding of Nature and provides incredible insights supplementing experimental observations. Hitherto, a plethora of literature discusses the computational discoveries in studying glasses in which positional disorder is inherent in their configurations. Motivated by active research and knowledge sharing, we developed a data analysis library in Python for computational materials science research. We hope to help promote scientific progress and narrow some technical gaps for the wide communities. The toolkit mainly focuses on physical analyses of glassy properties from the open-source simulator LAMMPS. Nevertheless, the code design renders high flexibility, with functionalities extendable to other computational tools. The library provides data-driven insights for different subjects and can be incorporated into advanced machine-learning workflows. The scope of the data analysis methodologies applies not only to materials science but also to physics, chemistry, and beyond.
\end{abstract}

\maketitle

%\vspace{5mm}
\newpage
\tableofcontents

\newpage
%--------------------------------------main text----------------------------------
\section{Introduction}
\vspace{2mm}

Matters in Nature create fascinating phenomena~\cite{ball1997made}. Their diverse properties help build a colorful world. What is behind these is the physical rules of interatomic interactions. Unearthing the atomic-scale physical mechanisms of various materials and their corresponding properties has been one of the main research subjects in natural science. Back to ancient times, theoretical analysis under the condition of simplified hypotheses is the major way to tackle realistic problems. One elegant example is the Newton's law of motion
\begin{equation}
{\vec F} = m {\vec a}.
\label{ma}
\end{equation}
The 18th century also witnessed a lot of great advancements in the theories of different disciplines. With the fast development of industrial technologies at that time, experimental studies played a growing important role. The direct observations from advanced equipments enrich the micro-world and provide verifications to theoretical speculations. Nevertheless, the limited spatial-temporal resolution puts constraints on diving deeper. The later advent of computational facilities provides valuable opportunities for fast calculation in scientific research. This motivates the proposal of computer simulations to help understand the real world by modeling very simple systems. Since then, computer simulations have become more and more pivotal in daily research. Nowadays, with the accumulated knowledge and the rich data mining tools, computational studies provide considerable insights for various scientific puzzles and are turning into an independent discipline in some fields as well.

There are several types of computer simulations, differentiated by the time scale and length scale, in general. For example, first-principle calculations based on the density functional theory (DFT), molecular dynamics (MD) simulations based on Newton's law, and Monte-Carlo (MC) simulations. The underlying algorithms thus differ as well. 
One common feature of these simulations is the calculation product in the type of atomic configurations. The positions and other related information are recorded for post-processing. Starting from these configurations, we are able to characterize the quantities of interest for answering different questions. Therefore, there are two key steps in computational science: one is modeling and the other is data analysis. 
The former provides reasonable snapshots or trajectories from fundamental physical rules. The latter supports physics-driven data analysis. Both of them are critical in harnessing computational power.

Based on the atomic packing in the simulation box, a material is classified into two groups: one is crystal and the other is glass. 
In crystals, atoms occupy the lattice positions in a crystalline structure with long-range translational periodicity. The structure is overall well-defined but can live with various kinds of defects. The sampling space is small but the input structure is crucial. 
In glasses, the atomic packing is disordered, without any lattice and long-range order. The structure is ill-defined, which gives rise to an extraordinary sampling space. The initial input configuration is not important at all after sufficient equilibration and sampling. Thus, we are usually interested in the dynamical properties of glasses, in addition to the local atomic structure.

In the current work, we mainly focus on the classical MD simulations in which the interatomic interactions are predefined either analytically or empirically. There are several types of interactions available at this moment and will be introduced in detail below. In brief, we hold a number of particles inside of a simulation box and regulate their positions and velocities by the interatomic potential. Physical constraints are kept to map it to relevant experimental observations. While the simulation procedure can be complex and technical by itself, we are here developing the post-processing data analysis tool to mine the generated configurations in an efficient way.
The library is written in Python with advanced supporting libraries, especially Numpy~\cite{harris2020array} and Pandas~\cite{reback2020pandas,mckinney-proc-scipy-2010}. They are helpful in dealing with the data in the matrix, with homogeneous and heterogeneous data formats.
Even though the toolkit is applicable to both crystals and glasses, we adhere to the latter in the following examples based on our relevant research experiences. Some research case studies are available from Refs.~\cite{hu2015five,hu2022origin,hu2020physical,hu2022revealing,hu2018configuration,hu2023universality}
The source codes are designed and developed following industrial standards. This will benefit future automation design in other workflows, such as machine learning and robotics.

The manuscript is organized as follows. In Sec. II, we introduce some fundamentals of molecular dynamics simulations. The way how we design the code is provided in Sec. III. This is followed by detailed discussions on different aspects of data analysis in materials science, physics, and chemistry: structure (Sec. IV), dynamics (Sec. V), Hessian matrix (Sec. VI), and finally vector analysis (Sec. VII). We then make a brief summary in the conclusion section.

\vspace{5mm}
\section{Classical computer simulations}
\vspace{2mm}

The classical MD simulations are developed based on equation (\ref{ma}). There are many excellent textbooks to guide both beginners and experts. One example is "Understanding molecular simulation: from algorithms to applications" by Frenkel and Smit~\cite{frenkel2023understanding}. The other is "Computer simulations of liquids" by Allen and Tildesley~\cite{allen2017computer}. There are too many others to list. These two books are good starting points, especially for newcomers.

We briefly introduce the simulation methods here. To perform an MD simulation, we first need to create a box of an arbitrary shape (usually orthogonal) and generate a number of particles inside. Because of the limitation of the computational power, we always have to confine ourselves to a finite-size box with a total of $N$ ($\sim 10^4$) particles. The finite-size effect will always be a topic to discuss for scientific findings in computer simulations. But there are two important points to emphasize. One is the periodic boundary conditions (PBC) imposed along the chosen dimensions. PBC enables the continuation of particle interactions across the physical boundaries of the simulation box. By avoiding the direct interaction of one particle from its image, these minimal box lengths encourage us to consider the system as sufficiently large in simulations. Under such conditions, we can still increase the system size to address the finite-size effect. The best practice is to figure out the $N$-dependence of the physical variable. These considerations also tell how the physical mechanism will depend on the length scale.

Given proper temperature and pressure (or density), the initial state can be relaxed at different thermodynamic states. For example, to obtain a liquid state, the initial structure is going to melt at a temperature much higher than its melting temperature. After waiting for a long enough time, the system will be in an equilibrium state. Reasonable quantities like pressure, temperature, kinetic energy, and potential energy can be computed. This process can be repeated at different temperatures to investigate the properties of supercooled liquids~\cite{debenedetti2001supercooled}. In the running process, the particles' positions and velocities are updated based on their interactions within the neighborhood. There are several types of interactions currently, based on computation efficiency. 

\begin{figure}[htp] %[htp]i
\centering
\includegraphics[width = \textwidth]{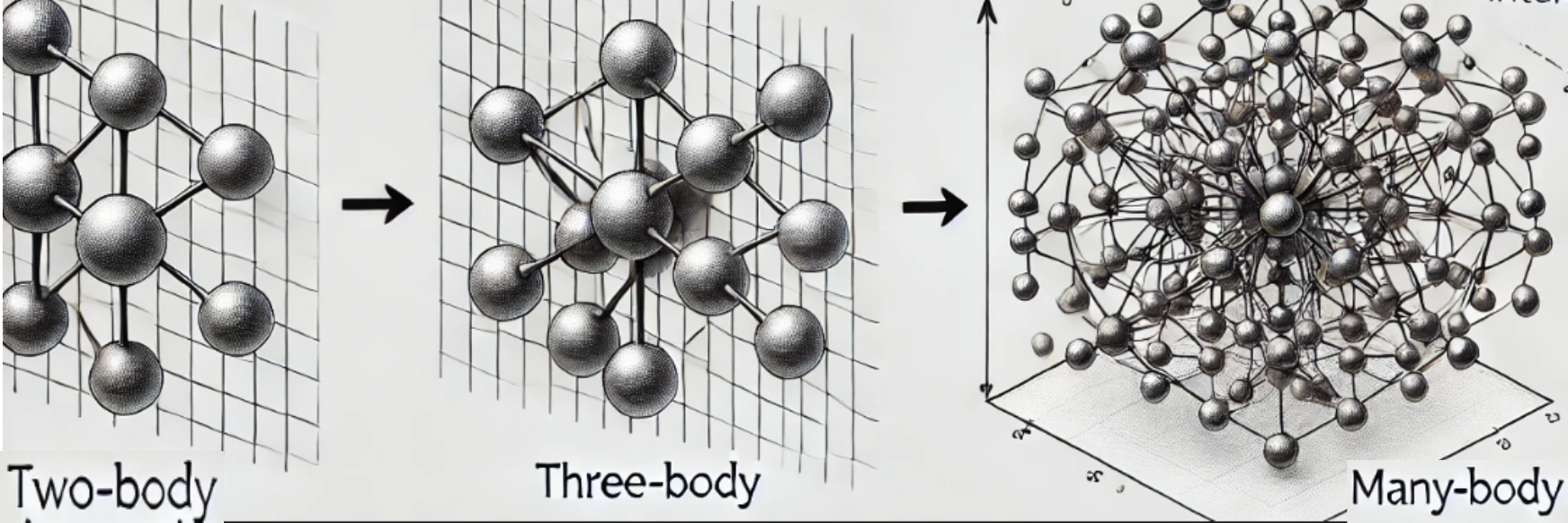}
\caption{
{\bf Illustration of the popular particle interactions in computer simulations}.
They are two-body (left), three-body (middle), and many-body (right) interactions.
For the pair potential, the interaction term only considers the pair distance and the properties of the involved particles. A typical example is the Lennard-Jones (LJ) interaction.
In the three-body situation, the angle formed by two bonds of the central particle is important for the interaction. A typical example is the Stillinger-Weber (SW) potential.
As for many-body interaction, the interaction between two particles includes the contribution from other atoms. A typical example is the empirical EAM potential.
The figure is generated by the ChatGPT-4o model (https://openai.com/index/learning-to-reason-with-llms/).
}
\label{fig1}
\end{figure}

Typically, they are two-body, three-body, and many-body interactions. Of course in Nature, the interatomic interactions should mainly be many-body. But the computational complexity will explode. For two-body interactions, calculating pair energy and force only involves two particles. The complexity is ${\mathcal O}(N^2)$. For the three-body interactions, such a calculation will include three particles to consider the bond angle. This induces $\mathcal{O} (N^3)$ complexity. Therefore, to include more realistic interactions, the computational complexity will grow extremely fast as $\mathcal{O} (N^n)$ for $n$-body interaction. 
In principle, if we extrapolate to many-body, the complexity would be $\mathcal{O} (N^N)$, which is infeasible for computations. In reality, to circumvent this difficulty, the empirical potential is usually used. A tabular data is first calculated to account for the contributions of the many-body terms as a function of pair distance. By implementing these pre-defined calculations, the many-body calculation is reduced to ${\mathcal O}(N^2)$ complexity, making it a pseudo-pair potential but with many-body inputs. This strategy is widely used in embedded-atom method (EAM) potentials, which are usually used to model metallic systems. Some illustrative plots are shown in Fig.~\ref{fig1} and more details of these interactions will be provided below.

%--------------------figure starts-------------------------------
%--------------------figure ends-------------------------------

Now we are going to introduce some interactions that are frequently used in MD simulations and are supportive of our library.

{\bf Two-body Interactions}. One of the most popular two-body interactions is the Lennard-Jones interaction~\cite{schwerdtfeger2024100}. Its potential energy is defined as
\begin{equation}
U_{\alpha \beta} (i, j) = 4 \epsilon_{\alpha \beta} \left[ \left(\frac{\sigma_{\alpha \beta}}{r_{ij}}\right)^{12} - \left(\frac{\sigma_{\alpha \beta}}{r_{ij}}\right)^6 \right]
 \label{LJ}
\end{equation}
for $r \leq r_{\rm c}$,
in which $\alpha$ and $\beta$ are particle species. $\epsilon$ and $\sigma$ are the corresponding pair cohesive energy and effective particle diameter. $r_{ij}$ is the pair distance between two particles $i$ and $j$ within a cutoff distance of $r_{\rm c}$. $U_{\alpha \beta} (i, j)$ is usually truncated and shifted to zero at $r_{\rm c}$, including its force derivative. Only the pair properties are included in the calculation. This potential form and its derivatives have been used widely in the research of glass physics. One of the authors widely tuned this potential to study the glass-forming ability of binary alloys~\cite{hu2023data,hu2022glass,hu2020glass,hu2019tuning}.

{\bf Three-body Interactions}. A typical example of three-body interaction is the Stillinger-Weber (sw) potential~\cite{stillinger1985computer}. The potential energy is the sum of a two-body term and a three-body term, defined as 
\begin{equation}
\begin{split} 
U(i, j) & =  \sum_i \sum_{j > i} \phi_2 (r_{ij}) +
        \sum_i \sum_{j \neq i} \sum_{k > j}
        \phi_3 (r_{ij}, r_{ik}, \theta_{ijk}) \\
\phi_2(r_{ij}) & =  A_{ij} \epsilon_{ij} \left[ B_{ij} (\frac{\sigma_{ij}}{r_{ij}})^{p_{ij}} -
                  (\frac{\sigma_{ij}}{r_{ij}})^{q_{ij}} \right]
                  \exp \left( \frac{\sigma_{ij}}{r_{ij} - a_{ij} \sigma_{ij}} \right) \\
\phi_3(r_{ij},r_{ik},\theta_{ijk}) & = \lambda_{ijk} \epsilon_{ijk} \left[ \cos \theta_{ijk} -
                  \cos \theta_{0ijk} \right]^2
                  \exp \left( \frac{\gamma_{ij} \sigma_{ij}}{r_{ij} - a_{ij} \sigma_{ij}} \right)
                  \exp \left( \frac{\gamma_{ik} \sigma_{ik}}{r_{ik} - a_{ik} \sigma_{ik}} \right)\end{split},
\label{sw}
\end{equation}
in which the free parameter set $\{ A, B, \epsilon, \sigma, p, q, a, \lambda, \theta_0 \}$ are useful to simulate different substances, for example, silicon and water~\cite{stillinger1985computer,molinero2009water}. Calculating the three-body term $\phi_3$ considers the angle of the two bonds connecting the center particle $\theta$. The angular constraint creates local anisotropy in the potential, such as the local tetrahedral order that is favored in water and silicon.

{\bf Many-body Interactions}. The quite often used many-body potential is the Embedded-atom Method (EAM) potential~\cite{finnis1984simple,daw1984embedded}, defined as
\begin{equation}
U_{\alpha \beta}(i, j) = F_\alpha \left(\sum_{j \neq i}\ \rho_\beta (r_{ij})\right) +
      \frac{1}{2} \sum_{j \neq i} \phi_{\alpha\beta} (r_{ij}),
\label{eam}
\end{equation}
in which $\phi$ is the two-body term and $F$ accounts for the many-body part by considering the combinatorial contributions of distance-dependent atomic electron density $\rho$. Therefore, the calculation of $U_{\alpha \beta}(i, j)$ is actually pairwise. This supports the simulation of realistic metallic materials. In addition, these terms are empirical by considering experimental measurements and DFT calculations. EAM potentials have been widely used in the field of metallic glasses, and other metallic materials.

Due to the absence of constraints on $\alpha$ and $\beta$, these potentials are ready to simulate multi-component systems. The extreme is the polydisperse system with the particle diameter sampled from a probability distribution like Gaussian. Therefore, computer simulations are rather flexible to supplement experimental studies in various fields. They also serve as benchmark tools for theoretical investigations.

Another important consideration is the timescale. The MD simulations run discretely by considering a series of timesteps. Each iteration between two consecutive steps accounts for a short period $\delta t$, usually 1 or 2 fs ($10^{-15}$ s). A large timestep is likely to cause unphysical movements. This fact poses an upper limit on the total simulation time that we can run. But they are perfect for some phenomena with fast dynamics, which are beyond the capability of experiments.
The usual length of the simulation time is on the scale of nanoseconds (ns) but can be extended to microseconds (ms) by GPU acceleration. Still, these long simulations may use months of computational time with multi-cores. This will limit our capability to study physical phenomena with slow dynamics. Some are called rare events. The prototypical example is crystallization, in which a phase will cross the free energy summit and transform into another phase. The barrier can be ultra-high, prohibiting us from observing some phenomena in the simulation timescale. This is especially true for protein folding-unfolding transition. Enhanced sampling techniques were born for these cases. In principle, a bias potential will be added to help fill in the potential energy well and help the system to get out of a state. In recent decades, enhanced sampling has been one of the most important methods for computational studies of phase transformations. The details are out of the scope of the current work, but available from the literature, such as ref.~\cite{laio2002escaping}.

With these processes and features in mind, the simulation usually requires an ensemble. This is critical for temperature and pressure control and provides a way for statistics. There are several ensembles available, such as the microcanonical ensemble (NVE), canonical ensemble (NVT), isobaric-isothermal ensemble (NPT), and grand-canonical ensemble ($\mu$VT). ($V$: volume, $P$: pressure, $T$: temperature, $\mu$: chemical potential). The ensemble is chosen case-by-case, depending on the scientific problem to study. During the simulations, a number of $M$ configurations are dumped for further data analysis, which falls into the main scope of the current library.

In short, classical MD simulations are the tool to generate configurations of a physical model with a regulation potential among particles to follow Newton's law of motion. By analyzing such configurations can we extract hidden insights to understand various scientific problems. Our library aims to provide an efficient framework for these analyses.
So far, many simulation softwares have been developed for different purposes. Some examples are the Large-scale Atomic/Molecular Massively Parallel Simulator (LAMMPS), GROMACS, and ASE.
Our library inputs the atomic configurations. Therefore, as long as the configuration information is readily read to the modules, the data analysis methods are applicable, no matter which software is used. A possible extension to the reader module may be necessary. This is true in the cases of DFT calculations and MC simulations. Therefore, our library is independent of simulators and simulation strategies. Even though it is designed mainly for classical MD simulations with LAMMPS as an example, it is easily extended to other use cases.

% by Jiachuan
\vspace{5mm}
\section{Code design strategies}
\vspace{2mm}
% data structure
% code structure
% CI/CD

\begin{figure}[htp] %[htp]i
\centering
\includegraphics[width = \textwidth]{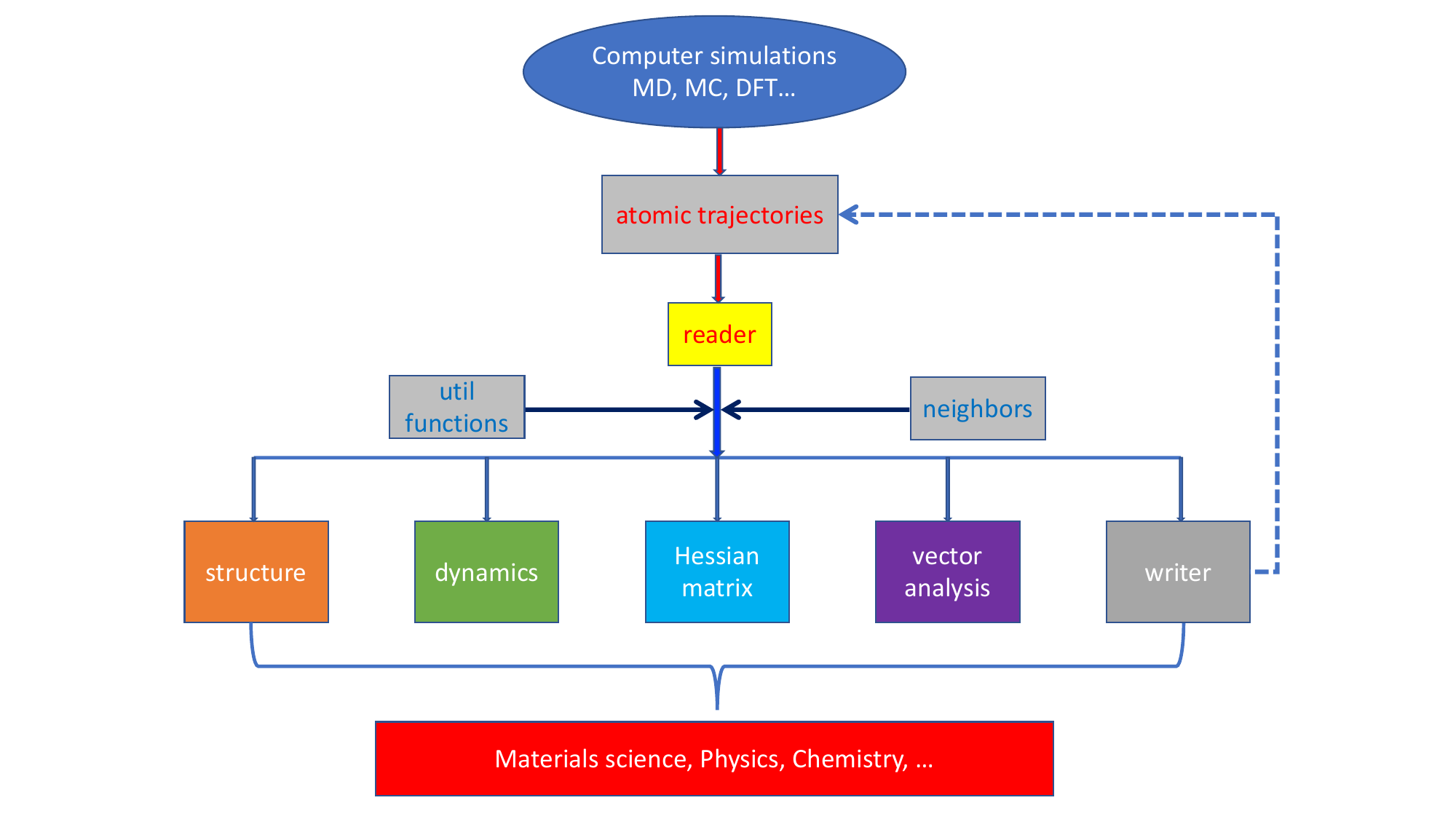}
\caption{
{\bf Flow chart of the high-level code design}.
}
\label{fig2}
\end{figure}

In this Python library, we first focus on some data analysis for disordered materials. However, the analysis itself is independent of the material system used.
After obtaining atomic trajectories, the analysis in the aspects of structure, dynamics, Hessian matrix, and vector analysis is ready to perform (see Fig.~\ref{fig2}). More details of each section will be discussed below.
The analysis will provide insights to investigate scientific problems in materials science, physics, chemistry, and others. We also plan to extend other functions to other systems with specific analysis in the future. 

%--------------------figure starts-------------------------------
%--------------------figure ends-------------------------------

In the section on structure analysis, we provide several functions on the pair correlation functions, structure factors, bond orientational orders, pair entropy, and other specific order parameters like local tetrahedral order for silicon- and water-like systems, and nematic order for anisotropic models and liquid crystals. Remarkably, the spatial correlation function and the Fourier transformation of an arbitrary order parameter are now available from the module. Such an order parameter can be a boolean, scalar, vector, or even tensor. Both float numbers and complex numbers are supported in the calculation. This provides flexibility for any predefined order parameter.

Two types of movements are considered when accounting for the dynamic properties. One is the absolute displacement field and the other is the coarse-grained version with cage-relative motion as a typical example. Based on the displacement field, both slow dynamics and dynamical heterogeneity can be evaluated. The four-point density-density correlation functions are included as well, which provides a way to estimate the high-order correlators. Another useful module is the time correlation of an arbitrary order parameter. This is helpful, for example, to calculate the velocity-velocity auto-correlation function and the lifetime of a certain order. The spatial-temporal correlation provided so far are important component in scientific research.

Then we will introduce how to measure the Hessian matrix of a model, which is the basic step to analyze the vibration and thermal properties of various materials. The Hessian matrix is determined by the interatomic interaction, learning of which will deepen the understanding of the MD simulation itself. Standing on the eigenvectors of the Hessian matrix, we introduce an independent section on the general analysis methods for a vector field. Both physics-driven quantities and general analysis are provided. High-dimensional vector or matrix analysis will be an important subject in the era of artificial intelligence. These basic analyses serve as a starting point for future deeper diving.

Practically, as shown in the flow chart of Fig.~\ref{fig2} the reader module will read the atomic trajectories generated by different computational methods. The neighbors module provides support to characterize the neighboring atoms for each particle. Some frequently used mathematical functions are encapsulated in util functions. These two modules will support the calculation of structural, dynamical, and other analyses from the atomic trajectories. In addition, a writer module will help convert the processed data to atomic trajectories, which may be helpful for data visualization. 

The above analyses are designed in separate modules, which provides flexibility to realize multi-functional implementations. The codes are integrated with the industrial standards. They may serve as a learning material for future software design.

\vspace{5mm}
\section{Structure}
\vspace{2mm}
The structure-property relationship is central to materials science. Building this correlation is the research focus for many decades. For crystals, a lot of achievements are available from the literature. However, this is far from mature for disordered materials. The lack of long-range ordering inhibits the definition of a sound order parameter to characterize the structure. Local atomic packings show enormous fluctuations in space and time. The non-equilibrium nature of glasses makes the sampling space enormous. There are ongoing efforts to characterize the structure of the disordered systems. Here we briefly introduce some structural analysis methods developed in the library.

\subsection{Pair correlation function}. 
The most popular measurement is from the pair correlation function (PCF) or radial distribution function (RDF). It considers the number density at a distance $r$ from a reference particle relative to the ideal gas, defined as
\begin{equation}
g(r) = \frac{1}{N \rho} \left< \sum_{i \neq j} \delta (r - |\vec r_{ij}|) \right>,
\label{pcf}
\end{equation}
where $\rho$ is the number density of the simulation system. In practice, this calculation runs over all pairs in the system and bins their pair distances into a histogram. This characterization applies to different atomic species, defined as 
\begin{equation}
g_{\alpha \beta}(r) = \frac{V}{N_{\alpha} N_{\beta}} \left< \sum_{i \neq j; i \in \alpha; j \in \beta} \delta (r - |\vec r_{ij}|) \right>.
\label{pcf_types}
\end{equation}
This calculation binarizes the pair distance based on their atomic species in addition to equation (\ref{pcf}). Both equations provide important information on the topological arrangements of atoms in space. These quantities, especially equation (\ref{pcf}), are related to the experimental measurements from the X-ray or neutron scattering. The two-body density correlation nature of $g(r)$ creates inherent limitations on the structural analysis for glasses, while it is crucial for crystals. It provides ensemble averaged spatial distribution information for a system. Thus important local information may be missed. As well, density is not always a proper order parameter for glasses. Note that the pair orientation is neglected in the calculations. Figuring out the key order parameter is the grand challenge in the studies of complex materials~\cite{steinhardt1983bond}.

This idea nevertheless brings additional flexibility to measure the spatial correlation of the particle-level properties, which can be a boolean, a scalar, a vector, or a tensor. The numbers of these data objects can be either float or complex or mixed. For example, if we name this physical quantity as $\mathcal A$, then the spatial correlation of $\mathcal A$ can be 
\begin{equation}
g_{\mathcal A}(r) = \frac{1}{N \rho} \left< \sum_{i \neq j} \delta (r - |\vec r_{ij}|) {\mathcal A}_i {\mathcal A}_j \right> = \left< {\mathcal A}(r) {\mathcal A}(0) \right>.
\label{gA}
\end{equation}
The normalization of $g_{\mathcal A}(r)$ by $g(r)$ gives the decay of the pair correlation. Note that $\mathcal A$ belongs to \{boolean, scalar, vector, tensor\}, which requires different processes of data manipulation. In this library, we consider direct multiplication of float numbers, dot-product of vectors, and trace of matrix multiplication. If a complex number shows up in $\mathcal A$, the complex conjugate will be considered for the multiplication and the real part will be the final product. There is no constraint on the length of the vector or the shape of the tensor. This is somehow consistent with the spirit of machine learning data processing. In glass research, the order parameter is usually scalar in the float or complex format. However, the complexity of glass structure naturally raises the question of how many dimensions the order parameter needs to well characterize the disordered structure. This is probably where supervised machine learning comes in.
More physically sophisticated multi-dimensional order parameters and their correlations in the high-dimensional space are important topics to unveil the glass mysteries.
Increasing the internal complexity may induce the emergence of unexpected findings as well.

\subsection{Structure factor}
Structure factor is a two-body correlation function to characterize the density correlation in the Fourier space. It is directly measurable in experiments by scattering techniques. It is related to PCF in real space. To calculate the structure factor, we first need to get $\rho(\vec q)$ by Fourier transformation of the number density:
\begin{equation}
\rho({\vec q}) = \sum_j^{N} \exp(-i {\vec q} \cdot {\vec r_j}),
\end{equation}
in which $\vec q$ is the wave vector and $\vec r_j$ is the positional vector of particle $j$. The wave vector of a three-dimensional system can be defined from 
\begin{equation}
\vec q = \left[\frac{2\pi}{L_x} n_x, \frac{2\pi}{L_y} n_y, \frac{2\pi}{L_z} n_z \right],
\label{qvector}
\end{equation}
where $L$ represents simulation box size at different dimensions and $n$ dictates an integer. A key note on $\vec q$ is that the lowest wavenumber $|\vec q|$ is determined by $L$. A large $L$ will be required to access the low $|\vec q|$ regime.
A separate module is developed to design $\vec q$ for different dimensions as a util function. This is a critical step for any characterization in $k$-space.

The structure factor is thus calculated from
\begin{equation}
S({\vec q}) = \frac{1}{N} \left< \rho({\vec q}) \rho({{-\vec q}}) \right>.
\label{sq}
\end{equation}
In practice, we can take the complex conjugate for the multiplication. From equation (\ref{sq}), both one-dimensional curves and two-dimensional scatters can be obtained for analysis. The former is enough for isotropic systems while the latter is better for anisotropic ones.

Similar to $g_{\mathcal A}(r)$, the calculation of $S(\vec q)$ can also be extended to any physical quantity $\mathcal A$ belonging to \{boolean, scalar, vector, tensor\}. In these cases, we first need to perform the Fourier transformation of $\mathcal A$ as 
\begin{equation}
\rho_\alpha ({\vec q}) = \frac{1}{\sqrt{N_\alpha}} \sum_i^{N_\alpha} {\mathcal A}_i \exp(-i {\vec q} \cdot {\vec r_i}),
\label{Afft}
\end{equation}
where $N_\alpha$ is the particle number under consideration based on $\mathcal A$. In this calculation, the original data format of $\mathcal A$ is preserved, being a scalar or vector, or tensor. But the component is now a complex number. This is an important way for other analyses (see below).
Their correlation is available from
\begin{equation}
S_{\alpha\beta}({\vec q}) = \left< \rho_\alpha({\vec q}) \rho_\beta({-\vec q})) \right>.
\label{sqA}
\end{equation}

For the calculations of PCFs and structure factors, our library now supports their partial counterparts up to five components. This will help the data analysis for high-entropy alloys where multiple species are sitting on the simple lattices.
The calculation also does not depend on dimensionality.

\subsection{Bond-orientational order parameters in two dimensions}
For two-dimensional (2D) systems, the hexatic order is usually favored by the entropic effect and the energetic effect as well. 
The basic bond orientational order parameter can be defined as
\begin{equation}
\varphi_l(j)=\frac{1}{N_j}\sum_{m=1}^{N_j}\exp({i l \theta_m}),
\label{phil}
\end{equation}
which considers the pair orientations of the central particle $j$ and its neighbors $m$ with respect to a reference axis. $l$ is an important parameter to determine the local particle arrangements. For example, $l=6$ will favor the hexatic order, which ideally has a bond angle of $60^\circ$. Other order parameters with $l \neq 6$ are also accessible in some simple LJ systems. This order parameter intrinsically is vectorial, which provides both the amplitude and phase of a complex number. 

The calculation in equation (\ref{phil}) considers the contribution from each neighbor equally. This is not necessarily true in model systems as multi-component and even polydisperse systems are usually considered. In addition, the unbalanced non-additive interactions will create local anisotropy as well. In these circumstances, a weight vector or matrix is required for adjustments, such as
\begin{equation}
\varphi_l(j)=\sum_{m=1}^{N_j}\frac{\ell_{jm}}{\sum |\ell_{jm}|} \exp({i l \theta_m}),
\label{philweight}
\end{equation}
in which $\ell_{jm}$ is the bond length between the pair $jm$. This can be calculated by Voronoi tessellation in the library. A proper weight matrix is key to identifying the order parameters for complicated 2D models.
To reduce the thermal noise, time average or spatial average is helpful~\cite{tanaka2010critical}. Our library supports these average strategies in the form of amplitude and the complex number itself.

The spatial coherence of these order parameters provides an estimation of their spatial correlation length $\xi_l$. This calculation actually maps to equation (\ref{gA}), and exemplified as
\begin{equation}
g_l(r)=\frac{L^2}{2\pi r \Delta r N(N-1)} \left< \sum_{j \neq k} \delta(\vec r - |\vec r_{jk}|). \varphi_l(j) \varphi_l^*(k) \right> 
\end{equation}
This quantity is powerful enough to measure the orientational order correlation in two-dimensional systems. For example, in a polydisperse LJ liquid, $g_6(r)/g(r)$ will decay fast but with a reducing speed at lower temperatures. This demonstrates a growing correlation length of hexatic order. For crystals favoring hexatic order, it does not decay, which helps to characterize the degree of crystallization.

\subsection{Bond-orientational order parameters in three dimensions}
For the three-dimensional (3D) models, the local structure is more complicated than that of 2D. The characterization of the spatial orientation of a particle with respect to its neighbors is by no means easy. Spherical harmonics is an efficient way. Based on the positional vector $\vec r_{ij}$, it is defined as
\begin{equation}
q_{{lm}}(i) = \sum_i^{N_i} {\mathcal X_{ij}} \cdot Y_{{lm}} (\theta(\vec r_{ij}), \phi(\vec r_{ij})),
\label{qlm}
\end{equation}
in which $Y_{lm}$ is the spherical harmonics that are available from scipy library~\cite{2020SciPy-NMeth}with $l \geq 0$. In our library, we provide hard-coded values for $0 \leq l \leq 10$ as util functions. The weight matrix element $\mathcal X_{ij}$ takes the contribution of each pair separately into consideration. For example, if $\mathcal X_{ij} = 1/N_i$ means that the neighbors are treated equally. While, if $\mathcal X_{ij} = S_j / \sum_j S_j$ where $S_j$ is the polygon face area of the Voronoi polyhedra of particle $i$, the contributions of each neighbor will depend on the topological property from Voronoi tessellation.

A coarse-grained version of $q_{lm}(i)$ over its neighbors is calculated as
\begin{equation}
Q_{lm}(i)=\frac {1}{N_i+1} \left(q_{lm}(i) + \sum_{j=1} ^{N_i} q_{lm}(j) \right),
\label{Qlm}
\end{equation}
which is important to exclude the non-extendable local structures. For example, in 3D, the icosahedron is likely favored by the entropic effect and it cannot extend to large length scales~\cite{leocmach2012roles}. While $q_{lm}$ is effective in characterizing crystal-like bond orientational orders, it has a high value for icosahedra as well. To remove this effect, $Q_{lm}$ is a more proper order parameter. This computation is supported internally. 

Therefore, we can analyze other derivative order parameters based on $q_{lm}$ and $Q_{lm}$ in the same way. Below we introduce these order parameters available from the library, but with only either as an illustration. They can be replaced by each other in the formula.
In principle, $q_{lm}$ and $Q_{lm}$ are particle-level vectors of complex numbers with a length of $2l+1$. Below we are dealing with the corresponding matrix processing. The first order parameter is the rotational invariants, which are defined as their norms:
\begin{equation}
q_l=\sqrt{\frac{4\pi}{2l+1} \sum_{m=-l}^{l} \left| q_{lm} \right|^2},
\end{equation}
which shows the degree of local orientational order in a scalar form. The second is the bond solidity parameter:
\begin{equation}
s(i,j)=\frac {\sum_{m=-l}^l q_{lm}(i) q_{lm}(j)}{\sqrt{\sum_{m=-l}^l \left| q_{lm}(i) \right|^2}
\sqrt{\sum_{m=-l}^l \left| q_{lm}(j) \right|^2}},
\end{equation}
which measures the spatial alignment of the complex vectors between a pair. This is very helpful in characterizing the crystallization process. The last one is defined as
\begin{equation}
\hat{w_l}=w_l \left(\sum_{m=-l}^l \left| q_{lm} \right|^2 \right) ^{-\frac {3}{2}},
\end{equation}
in which 
\begin{equation}
\begin{split}
w_{l} = \sum_{m_{1} + m_{2} + m_{3} = 0}^{}
\left(
\begin{matrix}
l & l & l \\
m_{1} & m_{2} & m_{3} \
\end{matrix}
\right)
q_{lm_{1}}q_{lm_{2}}q_{lm_{3}} 
\end{split},
\end{equation}
where
\begin{equation}
\begin{split}
\left(
\begin{matrix}
l & l & l \\
m_{1} & m_{2} & m_{3} \
\end{matrix}
\right)
\end{split}
\end{equation}
is the Wigner 3 $j-$symbols that are available from the Sympy library and are provided as a util function. It is useful to characterize icosahedral-like structures.

The spatial correlation of $q_{lm}$ and $Q_{lm}$ is similarly as in 2D to be a use case of $g_{\mathcal A}(r)$ where $\mathcal{A}$ is a vector of complex numbers. For example, one of them can be defined as
\begin{equation}
G_l(r)=\frac{4\pi}{2l+1} \frac{\sum_{ij} \sum_{m=-l}^l Q_{lm}(i) Q_{lm}^*(j) \delta(\vec r_{ij} - \vec r)} {\sum_{ij} \delta(\vec r_{ij} - \vec r)}.
\end{equation}

Based on the flexibility provided by $\mathcal{X}$, $l$, and coarse-graining, this calculation is very powerful in considering different cases. The combination of different parameters is sometimes very important. For example, to characterize BCC, FCC, and HCP crystals, combining different order parameters of $Q$ and $W$ is powerful~\cite{leocmach2012roles}.

\subsection{Local tetrahedral order}
In three-dimensional models with anisotropic interactions, such as water, silicon, and silica, they favor local tetrahedral order. As tetrahedron is the smallest unit in three dimensions with the densest packing feature, it is popular in various substances. For example, their appearance in some electrolytes can make hierarchical structures. However, their geometries and properties greatly differ in different materials. It is fair to say that it is its internal flexibility that facilitates the formation of various structures and the appearance of diverse functionalities. One traditional way to characterize local tetrahedral order is defined as

\begin{equation}
q_{\rm tetra} (i)=1-\frac{3}{8} \sum_{j=1}^3 \sum_{k=j+1}^4 \left(\cos \varphi_{jk}+\frac{1}{3} \right)^2,
\end{equation}
in which the calculation involves the angles formed by the bonds with central particle $i$.

\subsection{Pair entropy}
Entropy is an important concept in physics~\cite{wehrl1978general}. It plays a pivotal role in determining the states and properties of many materials. In general, entropy is made of two parts: vibrational entropy and configuration entropy. The latter is related to the number of independent configurations at a state. It is by no means easy to accurately compute configurational entropy in a straightforward manner even in computer simulations. 

An effective way to approximate entropy is from the pair contributions, see, for example, a study of supercooled liquids~\cite{tanaka2010critical}. Thus, the pair entropy $S_2$ can be defined from the particle-level PCFs:
\begin{equation}
\begin{split}
S_2^i = -2 \pi \rho k_B \int_0^{r_m} [g_m^i(r) \ln g_m^i(r) - g_m^i(r)+1] r^2 dr \quad (3D) \\
S_2^i = -\pi \rho k_B \int_0^{r_m} [g_m^i(r) \ln g_m^i(r) - g_m^i(r)+1] r dr \quad (2D)
\end{split}.
\end{equation}
It provides a way to measure the degree of disorder or order in the system.
Since $g^i_m(r)$ will saturate to 1.0 from a certain distance like $r_m$ for disordered materials, the calculation can be terminated at this distance to boost the efficiency. In this calculation, our library also used an interpolation method based on Gaussian distribution to evaluate $g^i_m(r)$:
\begin{equation}
\begin{split}
g_m^i(r) = \frac{1}{4 \pi \rho r^2} \sum_j \frac{1}{\sqrt {2 \pi \sigma^2}} e^{-(r-r_{ij})^2/(2 \sigma ^2)} \quad (3D) 
\\
g_m^i(r) = \frac{1}{2 \pi \rho r} \sum_j \frac{1}{\sqrt {2 \pi \sigma^2}} e^{-(r-r_{ij})^2/(2 \sigma ^2)} \quad (2D)
\end{split},
\end{equation}
which aims to overcome the large fluctuations from the discrete nature of particle arrangements in space. This method is sometimes called Gaussian blurring. It is very useful to analyze other particle-level quantities $\mathcal A$, especially for visualization. An independent module is developed as a util function.

So far the calculation of $S_2$ only concerns the translational information. A different version by considering the PCFs of bond orientations is also helpful to unveil the local structure~\cite{zheng2014structural,ingebrigtsen2018structural}. This point has long been overlooked in studying glass structure. Even though it is overall isotropic, local anisotropy can appear. This consideration will be implemented in the future.

\subsection{Nematic order}
In a model system with a preferred orientation, the nematic order parameter will be helpful in characterizing the local structure and the overall orientation~\cite{de1993physics}. Some model examples include but are not limited to spin liquids, patchy particles, liquid crystals, and magnetic materials. Classically, it is measured from a tensorial order parameter $\bf Q$, with its component defined as 
\begin{equation}
Q_{\alpha \beta}^i = \frac{d}{2} {\vec u}^i_{\alpha} {\vec u}^i_{\beta} - \delta_{\alpha \beta}/2,
\end{equation}
where $\vec u$ is the vector representation of local orientation, $\delta$ is Kronecker delta function, and \{$\alpha$, $\beta$\} are Cartesian coordinates. In some cases, for example, when $\vec u$ is uniformly unit vector, a coarse-graining over the first neighbor shell is helpful:
\begin{equation}
Q_{\rm CG}(i) = \frac{1}{1+N_i} \left( Q_{\alpha \beta}^i + \sum_{j}^{N_i} Q_{\alpha \beta}^j \right).
\end{equation}
Coarse-graining is an effective strategy to reduce local fluctuations, which somehow is consistent with Gaussian blurring, but in different ways. An independent module for coarse-graining given the neighbor list is provided as a util function in the library.

To quantify the local orientation in a scalar, an order parameter is usually defined as twice the largest eigenvalue of $\bf Q$. This calculation is not efficient for a large system, as diagonalizing the particle-level matrices is costly. An alternative is by
\begin{equation}
H_i = \sqrt{Tr[{\bf Q}^i \cdot {\bf Q}^i] \cdot \frac{d}{d-1}}.
\end{equation}
So far, in our library, the calculations of nematic order are limited to 2D. An extension to 3D is straightforward and planned.

\vspace{5mm}
\section{Dynamics}
\vspace{2mm}
One of the main features differentiating MD simulations from DFT calculations is the capability to characterize the dynamical properties of the modeled system. The {\it ab-initio} MD simulation is a supplement at the accuracy level of DFT. The general idea is to consider the correlation between configurations in a trajectory~\cite{berthier2011theoretical}. In an equilibrium state, this can be tricky in calculations. To ensure sufficient usage of the data, the moving average strategy is used. That is, we compare a configuration with all the other configurations that have not been compared. The key indicator is the time interval $\Delta t$ between the two configurations. The caution is to count the number of averages correctly for each $\Delta t$~\cite{allen2017computer}. This strategy requires the configurations to be dumped in a uniform $\Delta t$, which we term a "linear dump".
In this way, the time average is effective in improving the statistics.
In LAMMPS, one can dump the configurations in the log scale. That is, the short-time information is richer than from the linear dump. We term this way as "log dump". At a short timescale, the particle movements are small. Therefore, a large amount of average is usually not necessary. 
A good strategy to characterize the dynamics is by combining linear dump and log dump, from the developer's experiences. Both short-time and long-time information is recorded from a single run.
Both types of calculations are supported in the library.

\subsection{Dynamics of the displacement field}

Now we are ready to introduce the dynamical property characterizations supported by our library. As a first step, the particle-level movement is calculated as $\Delta {\vec r}_i(t) = {\vec r}_i(t) - {\vec r}_i(0)$ by removing the PBCs. In 2D simulations of supercooled liquids, the Mermin-Wagner effect is significant to create a strong finite-size effect in the estimation of the dynamics~\cite{flenner2015fundamental,shiba2016unveiling,illing2017mermin,vivek2017long}. Under these conditions, the cage-relative movement is considered:
\begin{equation}
\Delta \underline{\vec r_j} = \left[{\vec r_j} (t) - {\vec r_j}(0) \right] - \frac{1}{N_j}\sum_{i}^{N_j} \left[ {\vec r_i}(t) - {\vec r_i}(0) \right],
\label{cagemotion}
\end{equation}
which in essence subtracts the motion of a particle from its neighbors. This can serve as an alternative way to consider particle dynamics. From the developer's perspective, it is not necessarily limited to the cage-relative consideration for the above case. It is actually flexible by considering different coarse-graining levels. The hierarchical dynamics can be studied.
Therefore, there are always two ways to calculate the following dynamical quantities, by $\Delta {\vec r_j}$ or $\Delta \underline{\vec r_j}$.  We use one of them as an illustration.

The first quantity available is the intermediate scattering function (ISF) from the self-correlation:
\begin{equation}
F_s(q,t) = \frac{1}{N} \left< \sum_{j=1}^{N}\exp\lbrack i {\vec q} \cdot \Delta {\vec r_j}\rbrack\right>.
\label{isf}
\end{equation}
This calculation is carried out at a specific wave vector $\vec q$, which usually corresponds to the first peak location of the structure factor. This aims to quantify the particle movement over its diameter. Nevertheless, both the wavenumber and orientation dependence of ISF are useful in some studies. In this calculation, the same particle type is considered. This is what we mean by self-correlation. In principle, the calculation of ISF should be done by looping over all distinct pairs between two configurations. But this will increase the computation complexity to $\mathcal O (N^2)$ before the ensemble average. This is infeasible in practice. The simplified version is thus considered and it is sufficient to reveal some key dynamical information. In addition, this calculation shall require $\vec q$ over different directions. To boost the computation performance, only the directions along the Cartesian axes are considered. In glassy dynamics, $F_s(q,t)$ can be fitted to the Kohlrausch-Williams-Watts (KWW) function to reveal three messages~\cite{alvarez1993interconnection}: the Debye-Waller parameter to quantify the solidity, the structural relaxation time for overall dynamics, and the stretched exponent to measure the degree of dynamical heterogeneity.

An overlap function with similar functionality is defined as 
\begin{equation}
Q(t) = \frac{1}{N} \left< \sum_{j = 1}^{N}\omega\left( \left| \Delta {\vec r}_i(t) \right| \right) \right>,
\end{equation}
in which $\omega(x)$ is a step function that returns 1 if $x>a$ otherwise 0. The cutoff value $a$ is user-defined, which reflects the distance that a particle moves. Empirically, it is $0.3\sigma$ ($\sigma$: diameter). Like the role of $\vec q$, this choice depends on the dynamics of interest over what length scale. The fluctuations of the instantaneous values of $Q(t)$ gives its susceptibility~\cite{lavcevic2003spatially}:
\begin{equation}
\chi_{4}\left( t \right) = \frac{1}{N}\left( \left< {Q\left( t \right)}^{2} \right> - \left< Q\left( t \right) \right>^{2} \right),
\label{x4}
\end{equation}
in which the overlap function can be replaced with $F_s(q,t)$. The dynamical susceptibility $\chi_4(t)$ provides a way to characterize the time dependence of dynamical heterogeneity. Its hill-like shape gives a characteristic timescale $\tau_4$ to maximize the dynamic heterogeneity.

Another important parameter for dynamics is the diffusion rate, which can be measured in experiments as well. In simulations, it is usually measured from the mean-squared displacements (msd):
\begin{equation}
\left< \Delta {r^2}(t)\right> = \frac{1}{N}  \left< \sum_{j=1}^{N} |\Delta {\vec r}(i)| ^2  \right>,
\label{msd}
\end{equation}
which can be used to help calculate the non-Gaussian parameter:
\begin{equation}
\alpha_{2}\left( t \right) = \frac{3\left< \Delta r^{4}\left( t \right) \right>}{5\left< \Delta r^{2}\left( t \right) \right>^{2}} - 1\left( 3D \right); 
\qquad
\alpha_{2}\left( t \right) = \frac{\left< \Delta r^{4}\left( t \right) \right>}{2\left< \Delta r^{2}\left( t \right) \right>^{2}} - 1\left( 2D \right).
\end{equation}
$\alpha_2(t)$ provides another way to quantify the dynamical heterogeneity.

An important concept in glass physics and critical phenomena is the correlation length~\cite{tanaka2010critical}. It measures the extent to which the particle motions are cooperative. The cooperative rearrangements will cost more energy than from the particle level. An effective method is the so-called four-point dynamical structure factor~\cite{lavcevic2003spatially}:
\begin{equation}
S_4\left(q,t \right) = \frac{1}{N} \left< W\left({\vec q},t \right)W\left( - {\vec q},t \right) \right>,
\end{equation}  
where
\begin{equation}
W\left( {\vec q},t \right) = \sum_{j = 1}^{N}{\exp\left\lbrack {\vec q} \cdot {\vec r}_{j}\left( 0 \right) \right\rbrack\omega\left( \left| \Delta {\vec r}_j(t) \right| \right)},
\end{equation}
in which $\vec q$ is the wave vector and $q$ is the corresponding wavenumber. The low-$q$ regime is valuable to characterize the dynamical correlation length $\xi_4$. This calculation will require a large system size to reach lower $q$ information.
The principle for $S_4$ is considering the structure factor of certain particles based on a predefined condition. Therefore, it is easy to extend to other properties $\mathcal A$ for a static correlation length.

\subsection{Time correlation of order parameters}
The other type of dynamics is the time correlation of order parameters $\mathcal A$. As above, the order parameter can be a scalar, vector, or tensor. The number can be either float or complex as well. In general, it is calculated as 
\begin{equation}
C_{\mathcal A}(t) = \left< \frac{{\mathcal A}(t) \cdot {\mathcal A}(0)}{{\mathcal A}(0) \cdot {\mathcal A}(0)} \right>,
\label{cA}
\end{equation}
for which certain normalization may be required for better data representation. When a complex number exists in $\mathcal A$, the complex conjugate is used in the calculation, and the real part is taken as the final product. This correlation function can reveal the lifetime of certain order metrics.

Here we provide several examples for this analysis in MD simulations. 
(i) If $\mathcal A$ is the velocity field, the velocity-velocity correlation function is a way to characterize the density of state from the thermal state directly.
(ii) If $\mathcal A$ is the bond orientational order $Q_6$ from equation (\ref{Qlm}), the lifetime of crystal-like order is measurable.

\vspace{5mm}
\section{Hessian Matrix}
\vspace{2mm}
As introduced above, the most important model-specific input is the interatomic potential. It determines the potential energy and force for each pair and thus governs the underlying model properties. To a deeper level, the second derivatives of the potential energy with respect to positional vectors give the Hessian matrix $\bf D$. Specifically, each element is defined as 
\begin{equation}
{D}_{ij}=\frac{1}{\sqrt{{m}_{i}{m}_{j}}}\frac{{\partial }^{2}U({x}_{1},{y}_{1},...,{x}_{N},{y}_{N})}{\partial {R}_{i}\partial {R}_{j}},
\label{hessian}
\end{equation}
in which $m$ is the particle mass, and $R$ is the coordinate (see ref.~\cite{hu2022origin}) for example. By diagonalizing $\bf D$, the eigenmodes with a pair of eigenfrequency and eigenvector are obtained. The number of modes will depend on the degree of freedom of the system. The appearance of a negative eigenvalue indicates the instability of the model system. To ensure the positive eigenvalues, the system is usually first minimized to the local minima by some algorithms~\cite{plimpton_fast_1995}, such as gradient descent and FIRE. In another world, the system is mapped to its inherent structure at $T=0$ from the thermal state. The main focus in glass research refers to the inherent state, even though the instantaneous configuration can also convey some messages.

So $\bf D$ reveals critical information for the stability of the system and the next-step motion. Therefore, analyzing the Hessian matrix is an important tool to study various material properties, especially the vibrations and the associated thermal properties. In a perfect single-crystal, the vibration mode at the local minima of the potential energy landscape is phonon. The atoms participate in the $\vec q$-dependent waves, following the dispersion relation $\omega = c q$ ($\omega$: wave frequency). To characterize the Hessian matrix, there are two ways. One is analytical and the other numerical.

The analytical method is feasible for simple models, for which the potential energy has an explicit formula with respect to the model parameters. A typical example is the LJ interaction. In our library, in addition to LJ, there are two types of pair interactions supported. One is the inverse power law potential:
\begin{equation}
 U(r) = A \epsilon \left( \frac{\sigma}{r} \right)^n,
 \end{equation}
 where $A$, $\epsilon$, $\sigma$ and $n$ are model parameters. 
 The other is an even simpler system:
 \begin{equation}
 U(r) = \frac{\epsilon}{\alpha} \left(1 - \frac{r}{\sigma} \right)^\alpha,
 \end{equation}
 where $\epsilon$, $\alpha$, and $\sigma$ are model parameters. $\alpha=2$ gives the Harmonic potential and $\alpha=5/2$ makes a Hertz potential.
 
 In the code design, a shift parameter of boolean type is provided to choose whether to shift the potential energy and force to 0 at the cutoff by a linear method.
 
 For more complicated models, the numerical method is easier to implement. It is realized by the small displacements of the particles and calculates the induced force change. In principle, these two methods will provide consistent results. 
 
 To diagonalize the Hessian matrix, there are a couple of methods. The one we used in the library is from the Numpy package. We emphasize that the limitation is the calculation capability for $N<10^4$, with a higher tolerance for 2D systems. 
 
 After the matrix diagonalization, a number of eigenmodes are available. One of the important objects is the eigenfrequency. The vibration density of states is thus obtained by
\begin{equation}
D(\omega )=\frac{1}{dN-d}\mathop{\sum}\limits_{\lambda }\delta (\omega -{\omega }_{\lambda }).
\end{equation}
From the spectrum, a boson peak at low frequencies is always observed for disordered materials. This is one of the most fundamental research topics in glass science.

\vspace{5mm}
\section{Vector Analysis}
\vspace{2mm}
Another important product from the Hessian matrix is the eigenvector for each vibration mode. The analysis of this vector field is of paramount interest in research. In addition, there are other vector fields from the simulation data, for example, velocity, angular velocity, force, and so on. If the vector field is treated as an order parameter, the above analysis will also be applicable. There are two types of such analysis, one is static and the other dynamical.
The authors believe that vector analysis will play a more critical role in scientific research in the future, especially ignited by machine learning techniques. We provide some simple methods in the library currently and hope to include more in the future.

\subsection{Static properties}
For a given vector field, one important feature is how each member contributes to the overall field. This is usually quantified by the participation ratio:
\begin{equation}
PR_n = \frac{\left[ \sum_i^N |\vec e_{i,n}|^2 \right]^2}
{N\sum_i ({\vec e_{i, n}} \cdot {\vec e_{i,n}})^2}.
\end{equation}
For a perfect phonon, the participation ratio is around 2/3. But in glasses, there exist plenty of quasi-localized modes, in which some particles contribute more than others. This is an important feature in determining the vibration and thermal properties of glasses. There is still a lot of ongoing research interest in this subject~\cite{hu2022origin,hu2023universality}.

The local vector alignment order parameter, defined as 
\begin{equation}
\Psi_i = \frac{1}{CN_i} \sum_j^{CN_i} {\vec e_i} \cdot {\vec e_j},
\end{equation}
quantifies the coherency of the vector of the central particle $i$ with respect to its neighbors $j$. This is helpful in pinpointing the local disturbance and vertex appearance in the vector field.

Another parameter is the phase quotient of the vector field, calculated by 
\begin{equation}
PQ_n = \frac{\sum_i^{N} \sum_j^{CN_i} {\vec e_{i, n}} \cdot {\vec e_{j, n}}}{\sum_i^{N} \sum_j^{CN_i} |{\vec e_{i, n}} \cdot {\vec e_{j, n}}|},
\end{equation}
in which the local coherency is considered. The upper bound $PQ_n=1$ means the acoustic nature of the mode, while the lower bound of -1 means the optical nature of the mode. If $PQ_n>0$, the mode is acoustic-like otherwise optical-like.

An important feature of a vector field is its divergence and curl. Divergence is scalar over all dimensions. Curl only exists in 3D as a vector. Differently, curl can only be defined in 3D as a vector, its direction shows the local rotational direction while its magnitude means the amplitude of such motion. Both divergence and curl can be defined anywhere in space within the vector field. For example:
\begin{equation}
\begin{split}
{\rm div} \ {\vec u_i} = \nabla \cdot {\vec u} = \frac{1}{N_i} \sum_j^{N_i} (\vec R_j - \vec R_i) \cdot (\vec u_j - \vec u_i)
\\
{\rm curl}\ {\vec u_i} = \nabla \times {\vec u} = \frac{1}{N_i} \sum_j^{N_i} (\vec R_j - \vec R_i) \times (\vec u_j - \vec u_i)
\end{split},
\end{equation}
in which $\vec R$ is the position vector, $\vec u$ is the vector field (e.g. eigenvector), and $N_i$ is the number of nearest neighbors of a particle.

The last one is the so-called vibrability defined as
\begin{equation}
\Psi_i = \sum_{l=1}^{dN-d} \frac{1}{\omega_l^2} |{\vec e}_{l, i}|^2,
\end{equation}
which calculates the susceptibility of particle motion to infinitesimal thermal excitation in the zero temperature limit~\cite{tong2014order}. In this equation, the full spectrum of vibration modes is taken into consideration. However, it is easy to extend the calculation for a portion of modes.

\subsection{Vector decomposition in real space}
For a vector field, it intrinsically has two components: transverse (T) and longitudinal (L). To decompose a vector field in computer simulations, an effective way is by constructing a Voronoi volume matrix $A$:
\begin{equation}
{A}_{i,j\alpha }=\frac{1}{{V}_{i}}\frac{\partial {V}_{i}}{\partial {R}_{j\alpha }},
\end{equation}
which accounts for how each particle’s displacement in a direction will change others’ local volume~\cite{beltukov2015transverse}. $V_i$ is the Voronoi volume of particle $i$ and $R_{j\alpha}$ is the $j$th particle position along dimension $\alpha$ ($\alpha \in \{x, y, z\}$) in the inherent structure. So, $A$ is an $N \times dN$ matrix. An eigenvector $e_\lambda$ is projected onto the transverse and longitudinal direction as
\begin{equation}
\begin{split}
{{{{\bf{e}}}}}_{\lambda ,{{{\rm{L}}}}}={A}^{\mathrm{T}}{(A{A}^{\mathrm{T}})}^{-1}A{{{{\bf{e}}}}}_{\lambda },
\\
{{{{\bf{e}}}}}_{\lambda ,{{{\rm{T}}}}}={{{{\bf{e}}}}}_{\lambda }-{{{{\bf{e}}}}}_{\lambda ,{{{\rm{L}}}}}.
\end{split}.
\end{equation}
The transverse motion does not incur volume change while that is not true for the longitudinal motion. Based on the vector decomposition, the static property analysis methods introduced above apply to these components. One example is the corresponding density of states
\begin{equation}
{D}^{{{{\rm{x}}}}}(\omega )=\frac{1}{dN-d}\mathop{\sum}\limits_{\lambda }\frac{| | {{{{\bf{e}}}}}_{\lambda ,{{{\rm{x}}}}}| {| }^{2}}{| | {{{{\bf{e}}}}}_{\lambda }| {| }^{2}}\delta (\omega -{\omega }_{\lambda }),
\end{equation}
in which $x$ is T or L~\cite{hu2022origin}.

\subsection{Dynamical structure factors associated vector decomposition}
In addition to the vector decomposition in real space, the vector field can also be decomposed in reciprocal space based on the direction of the Fourier transformation of the vector field and the unit wave vector. First, we can take the Fourier transformation of a vector field by 
\begin{equation}
\tilde e = \sum_{i=1}^N \frac{\bf{e}_i^{\lambda}}{\sqrt{m_i}} \cdot \exp(i \bf{q} \cdot \bf{r}_i),
\end{equation}
which is in the same spirit as equation (\ref{Afft}). Then the longitudinal component can be calculated from 
\begin{equation}
E_{\lambda, L} (\bf{q}) = \frac{1}{N} \left| \hat{\bf{q}} \cdot \tilde e \right|^2.
\end{equation}
And the transverse component is available from 
\begin{equation}
E_{\lambda, T} (\bf{q}) = \frac{1}{N} \left| \hat{\bf{q}} \times \tilde e \right|^2.
\end{equation}
Starting from these, the dynamical structure factors of different components can be computed from
\begin{equation}
S_x (q, \omega) = \frac{k_B T}{M} \left(\frac{q}{\omega} \right)^2 \sum_{\lambda=1}^{dN-d} E_{\lambda, x} (\mathbf{q}) \delta(\omega - \omega_{\lambda}),
\end{equation}
in which $x$ is T or L. In these calculations, $\bf \hat q = \bf q/|\bf q|$ is the unit wave vector. The average mass is $M^{-1} = \sum_l m_l^{-1} /N_l$. 
Dynamical structure factors are important tools to understand the vibration properties of glasses~\cite{hu2022origin,hu2023universality}.

Another method to measure the dynamical structure factors is by using the auto-correlation function of velocity in the reciprocal space, i.e. the current-current correlation function~\cite{gelin2016anomalous}. In this strategy, we first transform the velocity field by
\begin{equation}
\vec j (\vec q, t) = \sum_{m=1}^N \vec v_m(t) \exp[i \vec q \cdot \vec r_m(t)],
\end{equation}
which can be decomposed into the transverse and longitudinal components by
\begin{equation}
\vec j_L (\vec q, t) = (\vec j (\vec q, t) \cdot \mathbf{\hat q}) \cdot \mathbf{\hat q},
\end{equation}
and 
\begin{equation}
\vec j_T (\vec q, t) = \vec j (\vec q, t) - \vec j_L (\vec q, t).
\end{equation}
Recalling equation (\ref{cA}), its time correlation is 
\begin{equation}
C_x(q, t)= \frac{1}{N} \left< \vec j_x({\vec q}, t) \cdot \vec j_x(-{\vec q}, 0) \right>,
\end{equation}
where $x$ is T or L. There is a quantitative relationship between $S_x(q, \omega)$ and $C_x(q, t)$ with an additional Fourier transformation in the time domain.
Another associated quantity is the magnitude of these components,
\begin{equation}
E_x (q) = \left < \frac{1}{N} \left| \vec j_x(\vec q, t) \right|^2 \right>.
\end{equation}

\vspace{5mm}
\section{Conclusion}
\vspace{2mm}
In summary, we present a brief theoretical introduction to the functionalities of our Python data analysis library, PyMatterSim. It takes the atomic configurations as the input, independent of the simulation methods and tools. 
The current explanations are focused on glassy materials based on our past research experience but are not necessarily within this material class.
The analysis codes cover research interests in the structure, dynamics, Hessian matrix, and vector field analysis of the computational systems. The package with test examples is available from the GitHub repository (\blue{https://github.com/yuanchaohu/pymattersim}) and can be easily installed by the tool PIP.
The documentation has been published online (\blue{https://yuanchaohu.github.io/pymattersim/index.html}) for further explanations and instructions. 
The fast analysis provided by this library will help understand the calculation details, boost research efficiency, and enhance knowledge sharing. In addition, the analysis modules can serve as an important component in the future machine-learning pipeline. We expect that our efforts will be helpful in various research fields and subjects, not limited to materials science, physics, and chemistry.

%----------------additional part---------------------------
\subsection *{Acknowledgments}
We thank J. You and Y.B. Hu for their help in reviewing and testing some codes and documentation.

%\subsection *{Author contributions.}
%Y.C.H. initialized and supervised this project, and designed and implemented the original codes. J.T. guided code design and contributed to refactoring. All the authors contributed to implementing and reviewing the codes. Y.C.H. is responsible for the code and document deployments.

%---reference
%\bibliography{mainref}
%apsrev4-2.bst 2019-01-14 (MD) hand-edited version of apsrev4-1.bst
%Control: key (0)
%Control: author (8) initials jnrlst
%Control: editor formatted (1) identically to author
%Control: production of article title (0) allowed
%Control: page (0) single
%Control: year (1) truncated
%Control: production of eprint (0) enabled
%

\balance
\end{document}